\documentclass[12pt,notoc]{JHEP3} % 10pt is ignored!

\usepackage{epsfig,multicol,bbm,amsmath}

\newcommand{\ep}{\epsilon}

\newcommand\fverb{\setbox\fverbbox=\hbox\bgroup\verb}
\newcommand\fverbdo{\egroup\medskip\noindent%
            \fbox{\unhbox\fverbbox}\ }
\newcommand\fverbit{\egroup\item[\fbox{\unhbox\fverbbox}]}
\newbox\fverbbox

\newcommand{\nablaslash}{\not{\hbox{\kern-3pt $\nabla$}}}

\title{Three-dimensional topologically gauged $\mathcal N=6$ ABJM type theories}

\author{Xiaoyong Chu and Bengt E.W.~Nilsson%\\
\\\\
Fundamental Physics\\
Chalmers University of Technology\\
SE-412 96 G\"oteborg, Sweden\\

{\tt {\footnotesize xiaoyong@student.chalmers.se,
tfebn@chalmers.se}}}

\abstract{In this paper we construct the $\mathcal N=6$ conformal
supergravity in three dimensions from a set of Chern-Simons-like
terms one for each of the graviton, gravitino, and R-symmetry gauge
field and then couple this theory to the $\mathcal N=6$
superconformal ABJM theory. In a first step part of the coupled
Lagrangian for this topologically gauged ABJM theory is derived by
demanding that all terms of third and second order in covariant
derivatives cancel in the supersymmtry variation of the Lagrangian.
To achieve this the transformation rules of the two separate sectors
must be augmented by new terms. In a second step we analyze all
terms in $\delta L$ that are of first order in covariant
derivatives. The cancelation of these terms require additional
terms in the transformation rules as well as a number of new terms
in the Lagrangian.  As a final step we
check that all remaining terms in $\delta L$ which are bilinear in
fermions cancel which means  that the presented Lagrangian and
transformation rules constitute the complete answer.
In particular we find in the last step  new terms in the scalar potential
containing either one or no structure constant. The non-derivative higher fermion terms
  in $\delta L$ that have
not yet been completely analyzed are briefly discussed. }


\begin{document}

%\maketitle  IS IGNORED %%%%%%%%%%%

\setcounter{page}{2}

\section{Introduction}

Conformal field theories in three dimensions have recently
experienced a number of interesting developments. The perhaps most
unexpected and profound results are the actual construction of a
seemingly unique three-dimensional maximally ($\mathcal N=8$)
superconformal theory by Bagger and Lambert, and by Gustavsson (BLG)
\cite{Bagger:2006sk,Gustavsson:2007vu,Bagger:2007jr,Bagger:2007vi},
along with its $\mathcal N=6$ variant (ABJM) by the authors of
\cite{Aharony:2008ug, Benna:2008zy}.

In  \cite{Gran:2008NT} an attempt was made to couple the $\mathcal
N=8$ BLG theory to conformal supergravity. After presenting a
detailed derivation of pure  $\mathcal N=8$ conformal supergravity,
this work went on to take  a first step towards the construction of
a Lagrangian describing the coupling of this theory to the BLG
theory. By checking that the supersymmetry variation of the coupled
Lagrangian  vanishes for
 terms of third and second order in covariant derivatives a set of coupling terms were
 obtained. This procedure also led to  a number of new terms
(as compared to the uncoupled theories) in the supersymmetry
variation of the two spin one gauge fields that enter these two
theories, namely the ones corresponding to the $SO(8)$ R-symmetry and
BLG gauge symmetry. However, the rigidity of the BLG theory seems at
this point to prevent a straightforward continuation of this
construction. For a
brief discussion why such topologically coupled theories might be of interest
in connection with M-theory and AdS/CFT, see the introductory
section of \cite{Gran:2008NT}.

In this paper we demonstrate that these results can be rather easily
obtained also for the $\mathcal N=6$ case. In fact, we will also show that
one can carry this construction, without meeting any serious obstacles,
 all
the way giving in the end  the entire Lagrangian and transformation rules.
As explained in the Conclusions,
however, the proof of supersymmetry is not yet completed since some of
the non-derivative higher fermion terms in the variation of the Lagrangian remain
to be checked.  The paper is organized as follows. In section two we
summarize our results on the Lagrangian and
transformation rules for the coupled theory. We start the derivation of these results  in
section three by constructing the $\mathcal N=6$ conformal
supergravity theory, and then go on in section four to review the
ABJM matter sector. With these two ingredients at hand, in section
five we take the first step in the process of coupling these two
theories by carrying out the same analysis as
 in \cite{Gran:2008NT} where it was done in detail in the $\mathcal N=8$ case.
 This step amounts to checking the cancelation of all terms with two covariant
 derivatives in $\delta L$. To get this to work we are forced to add new terms
 to the  transformation rules of the R-symmetry
 and ABJM gauge fields.
 The following section, section six, contains the second step where all terms
 in $\delta L$ containing one covariant derivative are checked and seen to cancel.
 This step requires a number of new terms in both the Lagrangian and
 in the transformation rules, in particular we find a $U(1)$ gauge field to
 play a special role.
 The terms in $\delta L$ bilinear in fermions and without derivatives are then discussed in
 section seven and shown to cancel. This step brings in new six-scalar terms in the potential
 which have either one or no structure constant.
 The theory obtained at this point can be shown to be the full theory. However, the
 proof of supersymmetry is yet not completed in all details. The terms in $\delta L$ that have not
 been checked so far are discussed in a concluding section.
 These terms are all without derivatives and contain more than two
 fermions.

\section{The complete Lagrangian and transformation rules of topologically
gauged $\mathcal N=6$ ABJM theories: a summary}

In this section we state the final result of this paper, that is,
the complete Lagrangian and supersymmetry transformation rules. The
invariance under the $\mathcal N=6$ supersymmetries is checked in
the following sections for all terms in $\delta L$ containing
covariant derivatives, as well as for all non-derivative terms that
are bilinear in fermions (including the supersymmetry parameter). At
this point in the construction
we are able to conclude that the results obtained constitute  the complete theory.

\subsection{The ansatz for the Lagrangian and transformation rules}

We find that the Lagrangian is given by (with $A=\pm\sqrt2$)
\begin{eqnarray}
 L&=&L_{sugra}^{conf}+L_{ABJM}^{cov}+\tfrac{1}{2}\epsilon^{\mu\nu\rho}C_{\mu}
 \partial_{\nu}C_{\rho}\label{CS:no idex}\\
  &&+iAe\bar\chi_{\mu}^{BA}\gamma^{\nu}\gamma^{\mu}\Psi_{Aa}
       (\tilde D_{\nu}\bar Z^a_B-\tfrac{i}{2} A\bar\chi_{\nu BC}\Psi^{Ca})+c.c.\label{squared:1}\\
  &&+i \epsilon ^{\mu\nu\rho} (\bar\chi_{\mu}^{AC}\chi_{\nu BC}) Z^B_a
       \tilde D_{\rho} \bar Z^a_A+c.c.\label{squared:2}\\
 &&-iA(\bar f^{\mu AB}\gamma_{\mu}\Psi_{Aa}\bar Z_B^a+
 \bar f^{\mu}_{AB}\gamma_{\mu}\Psi^{Aa}Z^B_a)\label{squared:3}\\
 &&-\tfrac{e}{8}\tilde R \vert Z\vert^2+\frac{i}{2}\vert Z \vert^2\bar f^{\mu}_{AB}\chi_{\mu}^{AB}\label{squared:4}\\
 % &&-ieA f^{ab}{}_{cd}(\bar \chi_{\mu AB}\gamma^{\mu}\Psi^{Dd})Z^A_a Z^B_b \bar Z^c_D+c.c.\notag\\
 &&+2ieA f^{ab}{}_{cd}(\bar \chi_{\mu AB}\gamma^{\mu}\Psi^{d[B})Z^{D]}_a Z^A_b \bar Z^c_D+c.c.\label{hatted:B}\\
 &&-i\epsilon^{\mu\nu\rho}(\bar \chi_{\mu AB}\gamma_{\nu}\chi_{\rho}^{CD})(Z^A_a Z^B_b\bar Z^c_C \bar Z^d_D)f^{ab}{}_{cd}\notag\\
 &&+\frac{i}{4}\epsilon^{\mu\nu\rho}(\bar\chi_{\mu AB}\gamma_{\nu}\chi_{\rho}^{AB})
 (Z^C_a Z^D_b \bar Z^c_C \bar Z^d_D)f^{ab}{}_{cd}\label{hatted:C}\\
 &&-\frac{i}{16}e \epsilon^{ABCD}(\bar\Psi_{Aa}\Psi_{Bb})\bar Z^a_C\bar Z^b_D+c.c.\notag\\
 &&+\frac{i}{16} e(\bar\Psi_{Db}\Psi^{Db})\vert Z\vert^2-\frac{i}{4}e(\bar\Psi_{Db}\Psi^{Bb})\bar Z^a_B Z^D_a\notag\\
 &&+\frac{i}{8}e(\bar \Psi_{Db}\Psi^{Da})\bar Z^b_B Z^B_a
   +\frac{3i}{8}e (\bar \Psi_{Db}\Psi^{Ba})\bar Z^b_B Z^D_a \label{hatted:D}\\
 &&-\frac{i}{16}eA(\bar \chi_{\mu AB} \gamma^{\mu}\Psi^{Bb}) \vert Z\vert^2 Z_b^A
 -\frac{i}{4}eA(\bar \chi_{\mu AB} \gamma^{\mu}\Psi^{Db}) Z^A_a Z^B_b \bar Z^a_D+c.c\label{hatted:E}\\
 &&-\frac{i}{4}\epsilon^{\mu\nu\rho}(\bar \chi_{\nu AB} \gamma_{\rho}\chi_{\mu}^{CD}) Z^A_a Z^B_b \bar Z^a_C \bar Z^b_D
  +\frac{i}{64}\epsilon^{\mu\nu\rho}(\bar \chi_{\nu AB} \gamma_{\rho}\chi_{\mu}^{AB})\vert Z\vert^4 \label{hatted:F}\\
 &&+\frac{1}{8}ef^{ab}{}_{cd} \vert Z\vert^2 Z^C_a Z^D_b \bar Z^c_C \bar
          Z^d_D+\frac{1}{2}e f^{ab}{}_{cd}Z^B_a Z^C_b Z^D_e \bar Z^e_B \bar Z^c_C \bar Z^d_D\label{V:prm}\\
 &&+\frac{5}{12 \cdot 64}e\vert Z\vert^6 -\frac{1}{32}e \vert Z\vert^2 Z^A_b Z^C_a \bar Z^b_C \bar Z^a_A+
 \frac{1}{48}e Z^A_a Z^B_b Z^C_d \bar
          Z^b_A \bar Z^d_B \bar Z^a_C,\label{V:prmprm}
  \end{eqnarray}
  where $c.c.$ refers to complex conjugation of the expression on the line where it occurs.

This Lagrangian has some features that need to be clarified at this
point\footnote{For comments about the introduction of a
dimensionless gravitational coupling constant and levels, see 
the concluding section.}. The first one concerns
the ABJM Dirac term that after gauging will be written in the
self-conjugate way\footnote{The SU(4) indices are used to keep track of complex conjugation
while the bar indicates if the SO(2,1) spinor index has been raised or lowered with a charge conjugation matrix 
(which are never written out explicitly).}
\begin{eqnarray}
- \tfrac{ie}{2} \bar \Psi^{Aa} \gamma^\mu  \tilde D_\mu  \Psi_{Aa}-
\tfrac{ie}{2} \bar \Psi_{Aa} \gamma^\mu \tilde D_\mu  \Psi^{Aa}.
\end{eqnarray}
Secondly,
the covariant derivative used here is defined by
\begin{eqnarray}
\tilde D_{\mu}\psi^{Aa}=\partial_{\mu}\psi^{Aa}+\tfrac{1}{4}\tilde
\omega_{\mu \alpha\beta}\gamma^{\alpha\beta}\psi^{Aa}+B_{\mu
B}^A\psi^{Ba}+\tilde A_{\mu b}^a\psi^{Ab}+qC_{\mu}\psi^{Aa},
\end{eqnarray}
where attention should be paid to the presence of the last term. The
Chern-Simons term for this abelian gauge field is written explicitly
in the Lagrangian given above and the reason for
giving the matter fields a charge $q$ under this explicit $U(1)$
will become clear later when we explain how we obtain the
topologically gauged ABJM Lagrangian. We will then also see that
$q^2=\tfrac{1}{16}$.

The purpose of this paper is to show that the above Lagrangian is
$\mathcal N=6$ supersymmetric.  We have found that this is the case
if the fields transform as follows:
\begin{align}
 \delta e_{\mu}{}^{\alpha}&= i \bar\epsilon_{gAB}\gamma^{\alpha}\chi_{\mu}^{AB},\\
 \delta \chi_{\mu}^{AB}&=\tilde D_{\mu}\epsilon_{g}^{AB} ,\\
 \delta B_{\mu ~B}^{~A}&=\frac{i}{e}(\bar f^{\nu AC}\gamma_{\mu}\gamma_{\nu}\epsilon_{g BC}-
  \bar f^{\nu}_{BC}\gamma_{\mu}\gamma_{\nu}\epsilon_{g}^{AC}
             )\cr
   &+\tfrac{i}{4}(\bar\epsilon_{BD}\gamma_{\mu}\Psi^{a(D} Z^{A)}_a-\bar\epsilon^{AD}\gamma_{\mu}\Psi_{a(D} \bar Z_{B)}^a)
        %  -\tfrac{i}{4}(\bar\epsilon_{BD}\gamma_{\mu}\Psi^{a[D}Z_a^{A]}
         % -\bar\epsilon^{AD}\gamma_{\mu}\Psi_{a[D}\bar Z_{B]}^a)
         \cr
   &-\tfrac{i}{2}(\bar\epsilon^{AC}_g\chi_{\mu DC}Z^D_a\bar Z^a_B-\bar\epsilon_{g BC}\chi^{DC}_{\mu}Z^A_a\bar Z^a_D)\cr
      & +\tfrac{i}{8}\delta^A_B(\bar\epsilon^{EC}_g\chi_{\mu DC}-
       \bar\epsilon_{gDC}\chi^{EC}_{\mu})Z^D_a\bar Z^a_E
       \cr
   &+\tfrac{i}{8}(\bar\epsilon^{AD}_g\chi_{\mu BD}-\bar\epsilon_{gBD}\chi^{AD}_{\mu})\vert Z\vert^2,\\
 \delta Z^A_a &=i\bar\epsilon^{AB}\Psi_{Ba}, \\
 \delta\Psi_{Bd}&=\gamma^\mu\epsilon_{AB}(\tilde D_\mu Z^A_d
 -iA\bar\chi_{\mu}^{AD}\Psi_{Dd})\cr
   &+ f^{ab}{}_{cd} Z^C{}_a Z^D_b \bar Z_B^c \epsilon_{CD}-
    f^{ab}{}_{cd} Z^A_a Z^C_b \bar Z_C^c\epsilon_{AB}\cr
   &+\frac{1}{4}Z^C_c Z^D_d \bar Z_B^c \epsilon_{CD}+\frac{1}{16} \vert Z\vert^2
   Z^A_d\epsilon_{AB},\\
 \delta \tilde A_{\mu~d}^{~c}&=-i(\bar\epsilon_{AB}\gamma_{\mu}\Psi^{Aa}Z^B_b-
 \bar\epsilon^{AB}\gamma_{\mu}\Psi_{Ab}\bar Z_B^a)f^{bc}{}_{ad}\cr
    &-2i(\bar\epsilon_g^{AD}\chi_{\mu BD}-\bar\epsilon_{gBD}\chi_{\mu}^{AD})Z^B_b\bar Z^a_A
    f^{bc}{}_{ad},\\
    \delta C_{\mu}&=-iq(\bar\epsilon_{AB}\gamma_{\mu}\Psi^{Aa}Z^B_a-
 \bar\epsilon^{AB}\gamma_{\mu}\Psi_{Aa}\bar Z_B^a)\cr
    &-2iq(\bar\epsilon_g^{AD}\chi_{\mu BD}-\bar\epsilon_{gBD}\chi_{\mu}^{AD})Z^B_a\bar
    Z^a_A\,,
\end{align}
where $ \epsilon_m^{AB}=A\epsilon_g^{AB}=\epsilon^{AB}, A=\pm
\sqrt{2}$ and $q^2=\tfrac{1}{16}$.

Finally we note that explicit covariant derivatives appear only in two terms in the Lagrangian,
namely the supercurrent term and, on the following line in the Lagrangian
above, the $\chi\chi ZDZ$ term. There is also an explicit covariant derivative
in the transformation rules of the Rarita-Schwinger field and the ABJM fermion.
In this context we note that in the
latter case the derivative is made supercovariant by adding a
second term giving the factor $(DZ-\chi\Psi)$. The same has to be done in the
supercurrent term in the Lagrangian but with an extra  factor of $\tfrac{1}{2}$. Note, however, that
the other derivative term in $L$ is not augmented with a similar term. We
have checked that such a term, which would be cubic in $\chi$, has
zero coefficient. Thus all terms with more than two $\chi$ fields
are in fact absorbed into the covariant derivatives and field
strengths in the Lagrangian.

The demonstration of supersymmetry carried out in the following
sections is divided into several steps starting with a construction
of ${\mathcal N=6}$ conformal supergravity. This is  followed by
adding on the ABJM theory and a stepwise incorporation of various
subsets of the interaction terms given above as supersymmetry is
checked for more and more terms in $\delta L$, organized in
decreasing order in covariant derivatives.

\section{Pure topological ${\mathcal N=8}$ and ${\mathcal N=6}$ supergravity in three dimensions}

For $\mathcal N=1$ a conformal and locally supersymmetric gravity theory in three
dimensions consisting of two Chern-Simons like terms was
shown to exist by Deser and Kay in \cite{Deser:1982sw} using methods that are
generalized to ${\mathcal N=8}$ in  \cite{Gran:2008NT} and in this paper to six
supersymmetries.  In \cite{VanNieuwenhuizen:1985ff}  the $\mathcal N=1$  theory
was derived from the superconformal algebra
by  imposing constraints on some of the  field strength components, while
in  \cite{Lindstrom:1989SG} the same methods were used to obtain
a superconformal Lagrangian for any $\mathcal N$.

In \cite{Gran:2008NT} also the problem of coupling the ${\mathcal
N=8}$  conformal supergravity to the $\mathcal N=8$ BLG theory was
discussed and the Lagrangian partly derived. Here we will first
briefly review the construction of  ${\mathcal N=8}$ pure
topological supergravity as presented in \cite{Gran:2008NT}, and
then redo this for ${\mathcal N=6}$. The goal in the following
sections is then to derive the Lagrangian describing the
coupling of  this ${\mathcal N=6}$ topological gravity theory to the
ABJM theory where we in a first step follow \cite{Gran:2008NT}.

\subsection{$\mathcal N=8$ pure topological supergravity}

Following the work of Deser and Kay \cite{Deser:1982sw} the authors of \cite{Gran:2008NT}
constructed  the on-shell  Lagrangian of three-dimensional ${\mathcal N=8}$ conformal
supergravity using only the three gauge fields of 'spin' 2, 3/2 and 1, i.e.,
$e_{\mu}{}^{\alpha},\,\,\chi_{\mu}{},\,\,B_{\mu}^{ij}$.
The result is
\begin{eqnarray}
L&=&\frac{1}{2}\epsilon^{\mu\nu\rho}
Tr_{\alpha}(\tilde\omega_{\mu}\partial_{\nu}\tilde\omega_{\rho}+
\frac{2}{3}\tilde\omega_{\mu}\tilde\omega_{\nu}\tilde\omega_{\rho})
-\epsilon^{\mu\nu\rho}Tr_i
(B_{\mu}\partial_{\nu}B_{\rho}+\frac{2}{3}B_{\mu}B_{\nu}B_{\rho})\cr
&&-i e^{-1} \epsilon^{\alpha\mu\nu}\epsilon^{\beta\rho\sigma}(\tilde
D_{\mu}\bar{\chi}_{\nu}\gamma_{\beta}\gamma_{\alpha}\tilde
D_{\rho}\chi_{\sigma}),
\end{eqnarray}
which was in \cite{Gran:2008NT}  explicitly shown to have $\mathcal N=8$ supersymmetry under the following transformation rules of the dreibein and Rarita-Schwinger field:
\begin{equation}
\delta e_{\mu}{}^{\alpha}=i\bar\epsilon
\gamma^{\alpha}\chi_{\mu}, \,\,\, \delta\chi_{\mu}= \tilde
D_{\mu}\epsilon.
\end{equation}
By demanding supersymmetry for any value of the R-symmetry gauge field strength, one immediately concludes that  the gauge field must vary under supersymmetry as follows:
\begin{equation}
\delta B_{\mu}^{
ij}=-\frac{i}{2e}\bar\epsilon\Gamma^{ij}\gamma_{\nu}\gamma_{\mu}f^{\nu}.
\end{equation}

The covariant derivative appearing in the Lagrangian and in the
variation of the Rarita-Schwinger field takes the form
\begin{equation}
\tilde
D_{\mu}\epsilon=\partial_{\mu}\epsilon+\frac{1}{4}\tilde\omega_{\mu\alpha
\beta}\gamma^{\alpha \beta}\epsilon+ \frac{1}{4}B_{\mu}^{
ij}\Gamma^{ij}\epsilon,
\end{equation}
acting on a three-dimensional spinor in an $SO(8)$ spinor representation.

Thus we explicitly gauge both the $SO(1,2)$ Lorentz  and the $SO(8)$
R symmetry. Note that the spinors in the gravity sector, i.e., the
SUSY parameter and the Rarita-Schwinger field, are of the same
$SO(8)$ chirality while the spinor in the BLG theory is of opposite
chirality. The commutator of two supercovariant derivatives, acting
on an SO(8) spinor, is
\begin{equation}
[\tilde D_{\mu},\tilde D_{\nu}]=\frac{1}{4}\tilde
R_{\mu\nu\alpha\beta}\gamma^{\alpha\beta}+
 \frac{1}{4}G_{\mu\nu}^{ ij}\Gamma^{ij},
\end{equation}

It will be convenient to define  the dual R-symmetry and curvature fields
 \begin{equation}
G^{*\mu}_{ij}=\frac{1}{2}\ep^{\mu\nu\rho}G_{\nu \rho ij},\,\,\,
\tilde
R^{*\mu}{}_{\alpha\beta}=\frac{1}{2}\epsilon^{\mu\nu\rho}\tilde
R_{\nu\rho\alpha\beta}
\end{equation}
and similarly for $\tilde\omega$, as well as the double and triple
duals
\begin{equation}
\tilde
R^{**\mu,\alpha}=\frac{1}{2}\ep^{\alpha\beta\gamma}\tilde
R^{*\mu}{}_{\beta\gamma}, \,\,\, \tilde
R^{***}_{\mu}=\frac{1}{2}\ep_{\mu\nu\alpha}\tilde
R^{**\nu,\alpha}.
\end{equation}

 Also as in \cite{Deser:1982sw},  we define the spin 3/2 field strength 
\begin{equation}
f^{\mu}=\frac{1}{2}\ep^{\mu\nu\rho}\tilde D_{\nu}{\chi}_{\rho},
\end{equation}
which can be used to write the spin 3/2 conformal term in the Lagrangian as
\begin{equation}
-4i(e_{\mu}{}^{\alpha}e_{\nu}{}^{\beta}e^{-1})\bar f^{\mu}
\gamma_{\beta}\gamma_{\alpha} f^{\nu}.
\end{equation}

The standard procedure to obtain local supersymmetry is to start by
adding Rarita-Schwinger terms to the dreibein-compatible $\omega$ in
order  to obtain a supercovariant version of it. That is, we define
\begin{equation}
\tilde\omega_{\mu \alpha\beta}=\omega_{\mu \alpha\beta}+K_{\mu
\alpha \beta},
\end{equation}
where
\begin{equation}
\omega_{\mu\alpha\beta}=\frac{1}{2}(\Omega_{\mu\alpha\beta}-\Omega_{\alpha\beta\mu}+\Omega_{\beta\mu\alpha}),
\end{equation}
with
\begin{equation}
\Omega_{\mu\nu\alpha}=\partial_{\mu}e_{\nu}{}^{\alpha}-\partial_{\nu}e_{\mu}^{\alpha},
\end{equation}
and
\begin{equation}
K_{\mu\alpha\beta}=-\frac{i}{2}(\bar\chi_{\mu}\gamma_{\beta}\chi_{\alpha}-
\bar\chi_{\mu}\gamma_{\alpha}\chi_{\beta}-\bar\chi_{\alpha}\gamma_{\mu}\chi_{\beta}).
\end{equation}
This combination of spin connection and torsion is
supercovariant, i.e., derivatives on the supersymmetry parameter
cancel out if $\tilde\omega_{\mu \alpha\beta}$ is varied under the
ordinary transformations of the dreibein and Rarita-Schwinger field.

In  \cite{Gran:2008NT} the supersymmetry of the Lagrangian given
above for ${\mathcal N=8}$ conformal supergravity was demonstrated
in full detail which required a certain amount of Fierz
transformations. We will not discuss this further here. Instead we
turn to the $\mathcal N=6$ case and give some of the details in that
context.

\subsection{$\mathcal N=6$ pure topological supergravity}

Let us start from the fact that in the  ABJM theory
\cite{Aharony:2008ug} the supersymmetry parameter is written
$\ep_{AB}$, with two antisymmetric $SU(4)$ indices in the
fundamental representation, thus producing six complex components.
To get a parameter in the
 real six-dimensional  vector representation of $SU(4)=SO(6)$ we need to impose
 the complex self-duality condition 
 (recall that $\epsilon^{AB}=(\epsilon_{AB})^*$)\footnote{See the previous footnote.}
 \begin{eqnarray}
\epsilon^{AB}=\tfrac{1}{2}\ep^{ABCD}\epsilon_{CD}.
\end{eqnarray}
With these conventions the local supersymmetry transformations take the form
\begin{equation}
\delta e_{\mu}{}^{\alpha}=i\bar\epsilon^{AB}
\gamma^{\alpha}\chi_{\mu AB}, \,\,\, \delta\chi_{\mu AB}= \tilde
D_{\mu}\epsilon_{AB}.
\end{equation}

Our goal now is to find a conformal Lagrangian that is
supersymmetric  under the above $\mathcal N=6$ transformations of
the dreibein and the Rarita-Schwinger field together with a
transformation of the $SO(6)$ R-symmetry gauge field $B_{\mu
~B}^{~A}$ that will be determined in the course of the calculation.
This superconformal  $\mathcal N=6$ supergravity theory will then be
coupled to the ABJM theory in later sections.

As we will show below the Lagrangian is the same as for $\mathcal
N=8$ apart from the normalization of the R-symmetry Chern-Simons
term which differs by a factor of two. This is due to the fact that
the trace is over the fundamental $SU(4)$ representation (indices
$A,B,...$) instead of the vector representation as in the $\mathcal
N=8$ case. Thus we claim that the Lagrangian for $\mathcal N=6$ is
\begin{eqnarray}
L&=&\frac{1}{2}\ep^{\mu\nu\rho}
Tr_{\alpha}(\tilde\omega_{\mu}\partial_{\nu}\tilde\omega_{\rho}+
\frac{2}{3}\tilde\omega_{\mu}\tilde\omega_{\nu}\tilde\omega_{\rho})
-2\ep^{\mu\nu\rho}Tr_A
(B_{\mu}\partial_{\nu}B_{\rho}+\frac{2}{3}B_{\mu}B_{\nu}B_{\rho})\cr
&&-i e^{-1} \ep^{\alpha\mu\nu}\ep^{\beta\rho\sigma}(\tilde
D_{\mu}\bar{\chi}^{AB}_{\nu}\gamma_{\beta}\gamma_{\alpha}\tilde
D_{\rho}\chi_{\sigma AB}),
\end{eqnarray}
where the last term can also be written
\begin{eqnarray}
-4i(e_{\mu}{}^{\alpha}e_{\nu}{}^{\beta}e^{-1})\bar f^{\mu AB}
\gamma_{\beta}\gamma_{\alpha} f^{\nu}_{AB},
\end{eqnarray}
in terms of the Rarita-Schwinger field strength $ f^{\mu}_{AB}$
defined as in the $\mathcal N=8$ case discussed above.

The covariant derivative acting on for instance the susy parameter
is defined by
\begin{equation}
\tilde D_{\mu}\epsilon_{AB}=\partial_{\mu}\epsilon_{AB}+\tfrac{1}{4}\tilde\omega_{\mu\alpha\beta}
\gamma^{\alpha\beta}\epsilon_{AB}-B_{\mu}{}^C{}_A\epsilon_{CB}
-B_{\mu}{}^C{}_B\epsilon_{AC}.
\end{equation}
By demanding that terms proportional to the R-symmetry gauge field strength cancel among themselves
we find the following transformation rule for the $B_{\mu}$ field
\begin{equation}
\delta B_{\mu}{}^A{}_B=\tfrac{i}{e}(\bar f^{AC}_{\sigma}\gamma_{\mu}\gamma^{\sigma}\epsilon_{BC}-
\bar f_{BC}^{\sigma}\gamma_{\mu}\gamma_{\sigma}\epsilon^{AC}).
\end{equation}
This expression can also be written
\begin{equation}
\delta B_{\mu}{}^A{}_B=\tfrac{2i}{e}(\bar f^{AC}_{\sigma}\gamma_{\mu}\gamma^{\sigma}\epsilon_{BC}-
\tfrac{1}{4}\delta^A_B\bar f^{CD}_{\nu}\gamma_{\mu}\gamma^{\nu}\epsilon^{CD}),
\end{equation}
and hence is defined to be traceless (see comment at the end of this subsection).

The calculation now goes through exactly as for $\mathcal N=8$, using for instance expressions like
\begin{equation}
\delta\tilde\omega^*_{\mu,\nu}=-2i(\bar\epsilon^{AB}\gamma_{\mu}f_{\nu AB}-\tfrac{1}{2}g_{\mu\nu}
\bar\epsilon^{AB}\gamma^{\rho}f_{\rho AB})
\end{equation}
and leads to the following expression for  $\delta L$:
\begin{eqnarray}
\delta L&=& \tfrac{4}{e}\bar\epsilon^{AB}(\gamma_{\mu}\gamma_{\nu}f^{\mu}_{AB})\bar
f_{\sigma}^{ CD}\gamma^{\nu}\chi^{\sigma}_{ CD}\cr
&&+\tfrac{8}{e}\bar
f^{\mu}_{CD}(\gamma_{\nu}\gamma_{\alpha}f^{\nu CD})(\bar\epsilon_{AB}\gamma^{\alpha}\chi_{\mu}^
{ AB}
-\tfrac{1}{2}e_{\mu}{}^{\alpha}\bar\epsilon_{AB}\gamma^{\sigma}\chi_{\sigma}^{AB}),
\cr
&&+\tfrac{4}{e^2}(\bar
f^{\mu}_{AB}\gamma_{\nu}\gamma_{\mu})\gamma_{\gamma}\chi_{\rho}^{AB}\epsilon^{\nu\rho\sigma}
(\bar\epsilon^{CD}\gamma_{\sigma}f^{\gamma}_{CD}-\tfrac{1}{2}e_{\sigma}{}^{\gamma}
\bar\epsilon^{CD}\gamma_{\tau}f^{\tau}_{ CD}),\cr
&&-\tfrac{16}{e^2}(\bar
f^{\mu AB}\gamma_{\nu}\gamma_{\mu})\chi_{\sigma CB}\epsilon^{\nu\rho\sigma}\bar\epsilon_{AD}
(\gamma_{\tau}\gamma_{\rho}f^{\tau CD})\cr
&&+\tfrac{8}{e^2}(\bar
f^{\mu AB}\gamma_{\nu}\gamma_{\mu})\chi_{\sigma AB}\epsilon^{\nu\rho\sigma}\bar\epsilon_{CD}
(\gamma_{\tau}\gamma_{\rho}f^{\tau CD}).
\end{eqnarray}

As in the $\mathcal N=8$ case presented in detail in  \cite{Gran:2008NT}, to demonstrate  supersymmetry we need to Fierz this expression and show that it vanishes. However,
at this point we will diverge from the treatment of the $\mathcal N=8$ theory where
both the $SO(1,2)$ and the $SO(8)$ spinor indices were Fierzed together. Here we first Fierz only the
spacetime $SO(1,2)$ spinors and then instead apply representation theory arguments or alternatively cycling of  the $\mathcal N=6$ spinor  indices
to conclude the proof of supersymmetry.

The strategy is thus to use the three-dimensional Fierz identity
\begin{equation}
 \bar AB \bar C D = -\tfrac{1}{2}( \bar AD \bar C B+ \bar A\gamma^{\mu}D \bar C\gamma_{\mu} B)
 \end{equation}
 to write all terms in $\delta L$ above in a form similar to the second term, 
 i.e., with the two $f^{\mu}_{AB}$ in the
 same scalar
 factor. The result of this operation is a number of terms similar to the second term but
 with the $SU(4)$ indices appearing in various positions: The two $f^{\mu}_{AB}$ can have both indices
 contracted (as in the second term) as well as one (from the fourth term) or  no (from the remaining terms)
  contracted indices.

 To understand how these different terms are related to each  other it is convenient
 to recall from the appendix of ref.\cite{Gran:2008NT} that the terms in $\delta L$ can be Fierzed
into a combination of twelve mutually independent
expressions (disregarding for the moment the $SU(4)$
  indices) of the type
  $(\bar f_{\mu}  f_{\nu})(\bar \epsilon  \chi_{\rho})\ep^{\mu\nu\rho}$,
 $( \bar f_{\mu} \gamma^{\nu} f^{\mu})(\bar \epsilon  \chi_{\nu})$ etc.
%$(\bar f_{\mu}  f^{\mu})(\bar \epsilon \gamma^{\nu} \chi_{\nu})$,
%$(\bar f_{\mu}  f_{\nu})(\bar \epsilon  \gamma^{\nu} \chi^{\mu})$,
%$(\bar f_{\mu} \gamma^{\nu\mu} f_{\nu})(\bar \epsilon \gamma^{\rho}
%\chi_{\rho})$, and $(\bar f_{\mu} \gamma^{\nu\mu} f_{\rho})(\bar
%\epsilon \gamma_{\nu} \chi^{\rho})
 Then  considering  the fact that ${\bf 6}\times{\bf 6}={\bf 1}+{\bf 15}+{\bf 20}$ where, if written
 in terms of four fundamental indices,
  ${\bf 15}$ is antisymmetric and ${\bf 1}$ and ${\bf 20}$ are symmetric under an interchange of the two
  antisymmetric pairs of indices. Using these properties all terms in $\delta L$ with the expression
  $\bar f...f$ in any given representation can be collected
  and seen to cancel exactly.

  A second  way to obtain this result arises if we
consider the fact that the antisymmetrization of five SU(4) indices
vanishes. We can then relate all terms with different index
structures to the three independent ones ${\bf 1}$, ${\bf 15}$ and
${\bf 20}$, which can then be collected and seen to cancel
separately.

  This theory will only be supersymmetric if the gauged R-symmetry is $SU(4)$, i.e., trying
  to include an abelian factor does not work. This can be seen for instance by making use of the equation
  \begin{equation}
  \bar f^{AC}\chi_{BC}= - \bar f_{BC}\chi^{AC}+\tfrac{1}{2} \delta^A_B \bar f^{CD}\chi_{CD},
 \end{equation}
 that is a direct consequence of the self-duality properties of the two fields in the equation.
 Note that this particular combination of $f$ and $\chi$ appears for instance in the Chern-Simons
 term for the gravitino field where it is contracted with an R-symmetry gauge field. Demanding that
 this term in the Lagrangian is real implies, due to the second term on the right hand side above,
 that the $B_{\mu}{}^A{}_B$ field is traceless. The term that removes the trace is then responsible
 for the very last term in
 expression for $\delta L$ presented above, and is needed in order to conclude that all terms cancel.

Similarly to the SO(8) case, the theory considered here also has
local scale invariance (denoted by an index $\Delta$) and possesses
$\mathcal N=6$ superconformal (shift) symmetry (denoted by $S$) with
the following transformation rules (where $\phi$ is the local infinitesimal scale
parameter and $\eta$ the local shift parameter)
\begin{eqnarray}
\delta_{\Delta} e_{\mu}{}^{\alpha}&=&-\phi(x)e_{\mu}{}^{\alpha},\cr
\delta_{\Delta} \chi_{\mu}^{AB}&=&-\tfrac{1}{2}\phi(x)\chi^{AB}_{\mu},\cr
\delta_{\Delta} B_{\mu}^A{}_B&=&0,
\end{eqnarray}
and
\begin{eqnarray}
\delta_S e_{\mu}{}^{\alpha}&=& 0,\cr
 \delta_S \chi_{\mu}^{AB}&=&\gamma_{\mu} \eta^{AB}, \cr
 \delta_S B_{\mu~~C}^{~A}&=& -i(\bar \eta^{AB}
 \chi_{\mu BC}-\bar \chi^{AB}_{\mu}
 \eta_{BC}).
\end{eqnarray}

\section{The $\mathcal N=6$ ungauged ABJM theory}

In this section we review the (ungauged) superconformal matter
sector, i.e., the ordinary ABJM theory, to which we would like
to couple the superconformal gravity derived in the previous
section. The resulting "topologically gauged" ABJM theory is then
the subject of the following sections.

\subsection{Review of the ungauged $\mathcal N=6$ superconformal ABJM action}

The formulation of the $\mathcal N=6$ matter theory in
\cite{Aharony:2008ug} makes no reference at all to any three-algebra
structure constants in contrast to the situation for the $\mathcal
N=8$ BLG theory. However, as shown in \cite{Bagger:2008se} the ABJM
theory is easily rewritten in terms of such structure constants, a
fact that was further developed in \cite{Nilsson:2008vi} where the
theory was expressed in terms of an additional algebraic structure
related to generalized Jordan triple systems. This provides a new
interpretation of the index structure of the fields and the
structure constants in terms of an infinitely graded Lie
algebra\footnote{This algebra is further discussed in
\cite{Palmkvist:2009qq}.}. The particular  form of the ABJM action
that we find convenient to use here is presented in
\cite{Nilsson:2008vi}.

In this new form of ABJM action, the complex scalars and fermions
are defined to have the specific index structure $Z^{A}_{a}$ and
$\Psi_{Aa}$, while their complex conjugates have the index structure
$Z_{A}^{a}$ and $\Psi^{Aa}$. These fields are then
connected to a formulation of the theory where the structure
constants have two upper and two lower indices \cite{Nilsson:2008vi}. 
Furthermore, these
indices are antisymmetric in each pair separately
\begin{equation}
f^{ab}{}_{cd}=f^{[ab]}{}_{cd}=f^{ab}{}_{[cd]}\,.
\end{equation}

The action of the $\mathcal N$= 6 M2-theory can now be written as
follows:
\begin{eqnarray}
{\cal L} &=& -(D_\mu Z^{A}_a)(D^\mu \bar{Z}_A{}^a) - i \bar
\Psi^{Aa} \gamma^\mu
 D_\mu  \Psi_{Aa}
 \nonumber\\
 &&  - i f^{ab}{}_{cd}\bar {\Psi}^{Ad}
 \Psi_{Aa}Z^B_b \bar{Z}_{B}^{c}+
 2i f^{ab}{}_{cd}\bar {\Psi}^{Ad}  \Psi_{Ba}Z^B_b \bar{Z}_{A}^{c}
  \nonumber\\
 &&
 -\tfrac{i}{2}\ep_{ABCD} f^{ab}{}_{cd}\bar {\Psi}^{Ac}  \Psi^{Bd} Z^C_a Z^D_b
  -\tfrac{i}{2}\ep^{ABCD} f^{cd}{}_{ab}\bar {\Psi}_{Ac}  \Psi_{Bd}\bar{Z}_{C}^a \bar{Z}_{D}^{b}
\nonumber\\
\label{lagrange} && -V +\tfrac{1}{2}\ep^{\mu\nu\lambda}(
f^{ab}{}_{cd}A_{\mu}{}^d{}_{b}
\partial_\nu A_{\lambda}{}^c{}_{a}+ \tfrac{2}{3} f^{bd}{}_{gc} f^{gf}{}_{ae}
A_{\mu}{}^a{}_{b}  A_{\nu}{}^c{}_{d} A_{\lambda}{}^e{}_{f}) \,,
\end{eqnarray}
where the gauge fields $A_{\mu ~b}^{~a}$ naturally appear in the
covariant derivatives in the following form
\begin{equation} \label{treelva}
\tilde A_{\mu}{}^a{}_{b} =f^{ac}{}_{bd} A_{\mu}{}^d{}_{c}\,,
\end{equation}
and the potential takes the form
\begin{equation}
V = \tfrac{2}{3} \Upsilon^{CD}{}_{Bd}\bar\Upsilon_{CD}{}^{Bd} \,,
\end{equation}
\begin{equation}
  \Upsilon^{CD}{}_{Bd}= f^{ab}{}_{cd} Z^C_a{Z}^D{}_b \bar{Z}_B^c
  + f^{ab}{}_{cd}\delta^{[C}{}_B Z^{D]}_a{Z}^E_b \bar{Z}_{E}^{c} \,.
\end{equation}
The transformation rules for the six supersymmetries, parametrized
by the complex self-dual three-dimensional spinor $\epsilon_{AB}$,
read:
\begin{equation}
\delta Z^A_a=i \bar \epsilon^{AB}\Psi_{Ba}\,,
\end{equation}
\begin{equation}
\delta \Psi_{Bd}=\gamma^\mu D_\mu Z^A_d \epsilon_{AB} +
f^{ab}{}_{cd} Z^C_a Z^D_b \bar Z_B^c
 \epsilon_{CD}- f^{ab}{}_{cd} Z^A_a Z^C_b \bar Z_C^c \epsilon_{AB}\,,
\end{equation}
 \begin{equation}
\delta A_\mu{}^a{}_{b}=-i \bar \epsilon_{AB} \gamma_\mu \Psi^{Aa}
Z^B_b +
 i \bar \epsilon^{AB} \gamma_\mu \Psi_{Ab} \bar Z_{B}^{a}\,.
\end{equation}

This action can be shown to be $\mathcal N=6$ supersymmetric
provided that the structure constants obey the fundamental identity
\cite{Nilsson:2008vi} (see also \cite{Bagger:2008se})
\begin{equation} \label{fundid}
f^{a[b}{}_{dc} f^{e]d}{}_{gh} = f^{be}{}_{d[g} f^{ad}{}_{h]c}\,,
\end{equation}
and, under complex conjugation,
\begin{equation} \label{komplexkonjugering}
(f^{ab}{}_{cd})^\ast = f^{cd}{}_{ab} \equiv f_{ab}{}^{cd}.
\end{equation}

\section{Coupling $\mathcal N=6$ conformal supergravity to ABJM
matter: the result after cancelation of $(D_{\mu})^2$ terms in
$\delta L$}

In the two previous sections we have discussed both the ABJM theory
and $\mathcal N=6$ conformal supergravity, the latter derived
explicitly in section three.  The coupling of these two theories to
each other can be obtained in several ways. Here we will use the
method based on an expansion in powers of derivatives used
previously in ref. \cite{Gran:2008NT}. Thus, as the first step we
consider in this section  only the cancelation of terms in the
variation of the Lagrangian that are of second order in covariant
derivatives. Terms of third order in derivatives also appear but
only in the supergravity sector and have thus already been analyzed.
This procedure was demonstrated in \cite{Gran:2008NT} to produce
additional terms in the transformation rules for the spin one gauge
fields in addition to a set of coupling terms that render the theory
supersymmetric to this order in covariant derivatives. Applying this
strategy here we use the following terms as a starting point:
\begin{eqnarray}
L&=&L^{conf.}_{sugra}+L_{ABJM}^{cov.}+L_{supercurrent}^{cov},
\end{eqnarray}
where  $L^{conf.}_{sugra}$ has been given in a previous section, 
the covariantized ABJM Lagrangian
\begin{eqnarray}
L_{ABJM}^{cov.}&=&e( -(\tilde D_\mu Z^{A}_a)(\tilde D^\mu
\bar{Z}_A^a) - i \bar \Psi^{Aa} \gamma^\mu
 \tilde D_\mu  \Psi_{Aa}+L_{Yukawa}-V)+L_{CS(A)},\cr
 &&
 \end{eqnarray}
and 
 \begin{eqnarray}
L_{supercurrent}^{cov}=A ie (\bar \chi_{\mu AB}
\gamma^{\nu}\gamma^{\mu} \Psi^{B a})(\tilde
D_{\nu}Z^{A}_{a}-\tfrac{i}{2} \hat{A} \bar \chi_{\nu}^{AD}\Psi_{D a}
)+c.c.,
 \end{eqnarray}
 where the constants $A$ and $\hat A$ will be determined below.

The transformation rules at this point in the analysis are the ones
used in sections three and four but with fully covariant
derivatives, reproduced here for convenience,
\begin{eqnarray}
\delta e_{\mu}{}^{\alpha}&=&i\bar\epsilon^{AB}_g
\gamma^{\alpha}\chi_{\mu AB}\,,\cr \delta\chi^{\mu AB}&=& \tilde
D_{\mu}\epsilon^{AB}_g\,,\cr \delta
B_{\mu}{}^A{}_B&=&\tfrac{i}{e}(\bar
f^{AC}_{\sigma}\gamma_{\mu}\gamma^{\sigma}\epsilon_{gBC}- \bar
f_{BC}^{\sigma}\gamma_{\mu}\gamma_{\sigma}\epsilon^{AC}_g)\,,\cr
\delta Z^A_a&=&i \bar \epsilon^{AB}_m\Psi_{Ba}\,,\cr
 \delta \Psi_{Bd}&=&\gamma^\mu (\tilde
D_\mu Z^A_d -i \hat{A} \bar \chi_{\mu}^{AD}\Psi_{D d})\epsilon_{mAB}+
f^{ab}{}_{cd} Z^C_a Z^D_b \bar Z_B^c
 \epsilon_{mCD}- f^{ab}{}_{cd} Z^A_a Z^C_b \bar Z_C^c
 \epsilon_{mAB}\,,\cr
\delta A_\mu{}^a{}_{b}&=&-i \bar \epsilon_{mAB} \gamma_\mu \Psi^{Aa}
Z^B_b +
 i \bar \epsilon^{AB}_m \gamma_\mu \Psi_{Ab} \bar Z_{B}^{a}\,,
\end{eqnarray}
where the two (gravity and matter) supersymmetry parameters 
will be related below.

We will later need to add more  terms  in order to keep the theory
supersymmetric to the order of approximation we are then working.
Note, however, that the hatted coefficient $\hat{A}$ in the ansatz
is not determined by the $(\tilde D_{\mu})^2$ calculation below but
simply by demanding that the  $\tilde D_{\mu}Z$ factor in
$\delta\Psi$ be supercovariant, i.e., $\tilde D_{\mu}Z$ must be
replaced, as done in the ansatz, by $\tilde D_{\mu}Z-i\hat
A\bar\chi\Psi$ in order to eliminate terms where the derivative acts
on the supersymmetry parameter when this expression is varied. The
parameter $\hat A$ is then obtained as soon as the relation between
the ABJM and supergravity supersymmetry parameters are determined.
Note that the presence of a factor of $\tfrac{1}{2}$ in front of
$\hat A$ in the supercurrent term is common in supergravity and
follows from standard arguments. These features of the theory will
be verified in the next chapter when supersymmetry is implemented by
canceling terms in $\delta L$ with one derivative.

\subsection{Supersymmetry at order $(\tilde D_{\mu})^2$}

We start by performing the variation of the covariantized scalar and
spinor kinetic terms. The scalar one
\begin{equation}
L_1=-eg^{\mu\nu}(\tilde D_{\mu}Z^A_a)(\tilde D_{\nu}\bar Z_A^a),
\end{equation}
gives
\begin{eqnarray}
\delta L_1&=&2ie(\tilde D_{\mu}Z^A_a)(\tilde D_{\nu}\bar
Z_A^a)(\bar\epsilon^{BC}_g\gamma^{(\mu}\chi^{\nu)}_{BC}-
\tfrac{1}{2}g^{\mu\nu}\bar\epsilon^{BC}_g\gamma^{\rho} \chi_{\rho
BC})\cr &&+ie(\bar\epsilon_m^{AB}\Psi_{Ba}\tilde\Box\bar
Z_A^a+\tilde\Box
 Z^A_a\bar\epsilon_{mAB}\Psi^{Ba})\cr
 &&-eg^{\mu\nu}(-Z^A_b\delta \tilde A_{\mu}{}^b{}_a +\delta
 B_{\mu}{}^A{}_B Z^B_a)\tilde D_{\nu}\bar Z^a_A\cr
&&-eg^{\mu\nu}\tilde D_{\mu}Z_a^A(\delta \tilde A_{\nu}{}^a{}_b \bar
Z_A^b-\bar Z_B^a\delta
 B_{\nu}{}^B{}_A ).
\end{eqnarray}
Our first goal will be to cancel the first two lines. For the second
line we need the variation of the Dirac term
\begin{equation}
L_2=-iee_{\alpha}{}^{\mu}\bar\Psi^{Aa}\gamma^{\alpha}\tilde
D_{\mu}\Psi_ {Aa}.
\end{equation}
Its variation reads, after an integration by parts which produces a
torsion term $\tilde D_{\mu}e_{\alpha}{}^{\mu}=K_{\mu
\alpha}{}^{\mu}=\bar\chi_{\alpha}^{BC}\gamma^{\beta}\chi_{\beta
BC}$ (second line),
\begin{eqnarray}
\delta L_2&=&2e\bar\epsilon_g^{BC}\gamma^{\beta}\chi_{\rho
BC}e_{[\alpha}{}^{\mu}
e_{\beta]}{}^{\rho}\bar\Psi^{Aa}\gamma^{\alpha}\tilde
D_{\mu}\Psi_{Aa}\cr
&&+e(\bar\chi_{\alpha}^{BC}\gamma^{\beta}\chi_{\beta
BC})\bar\Psi_{Aa}\gamma^{\alpha}\delta\Psi^{Aa}\cr
&&-ie(\bar\Psi^{Aa}\gamma^{\mu}\tilde D_{\mu}\delta\Psi_
{Aa}+\bar\Psi_{Aa}\gamma^{\mu}\tilde D_{\mu}\delta\Psi^{Aa})\cr
&&-ie\bar\Psi^{Aa}\gamma^{\mu}(\tfrac{1}{4}\delta\tilde\omega_{\mu\alpha\beta}\gamma^{\alpha\beta}\Psi_{Aa}
+\delta\tilde A_{\mu a}{}^b\Psi_{Ab}+\delta B_{\mu A}{}^B\Psi_{Ba}).
\end{eqnarray}

Thus we need to compute $\gamma^{\nu}\tilde D_{\nu}\delta\Psi_
{Bd}$. We find, again using $\tilde D_{\nu}e_{\alpha}{}^{\mu}=K_{\nu
\alpha}{}^{\mu}$,
\begin{eqnarray}
\gamma^{\nu}\tilde D_{\nu}\delta\Psi_{Bd}&=&
(\gamma^{\nu}\gamma^{\mu}\tilde D_{\nu}\tilde D_{\mu}Z^A_d)\ep_{mAB}+
\gamma^{\nu}\gamma^{\mu}(\tilde D_{\nu}\ep_{mAB})(\tilde D_{\mu}Z^A_d)+
\gamma^{\nu}\gamma^{\alpha}\ep_{mAB}(\tilde D_{\mu}Z^A_d)K_{\nu\alpha}{}^{\mu}\cr
&&-i\hat A\gamma^{\nu}\gamma^{\alpha}\ep_{mAB}K_{\nu\alpha}{}^{\mu}(\bar\chi_{\mu}^{AD}\Psi_{Dd})
-i\hat A\gamma^{\nu}\gamma^{\mu}\tilde D_{\nu}\ep_{mAB}(\bar\chi_{\mu}^{AD}\Psi_{Dd})\cr
&&-i\hat A\gamma^{\nu}\gamma^{\mu}\ep_{mAB}(\tilde D_{\nu}\bar\chi_{\mu}^{AD}\Psi_{Dd}+
\bar\chi_{\mu}^{AD}\tilde D_{\nu}\Psi_{Dd})\cr
&&+\gamma^{\nu}\ep_{mCD}f^{ab}{}_{cd}(2(\tilde D_{\nu}Z^C_a)Z^D_b\bar Z^c_B+
Z^C_aZ^D_b(\tilde D_{\nu}\bar Z^c_B))\cr
&&-\gamma^{\nu}\ep_{mAB}f^{ab}{}_{cd}((\tilde D_{\nu}Z^A_a)Z^C_b\bar Z^c_C+Z^A_a
(\tilde D_{\nu}Z^C_b)\bar Z^c_C+Z^A_aZ^C_b(\tilde D_{\nu}\bar Z^c_C))\cr
&&+f^{ab}{}_{cd}(\gamma^{\nu}\tilde D_{\nu}\ep_{mCD}Z^C_aZ^D_b\bar Z^c_B-
\gamma^{\nu}\tilde D_{\nu}\ep_{mAB}Z^A_aZ^C_b\bar Z^c_C)\,.
\end{eqnarray}

Now since
\begin{eqnarray}
\gamma^{\nu}\gamma^{\mu}\tilde D_{\nu}\tilde D_{\mu}Z^A_d=\tilde\Box Z^A_d+\tfrac{1}{2}
\gamma^{\mu\nu}\tilde (F_{\mu\nu d}{}^e Z^A_e+G_{\mu\nu}{}^A{}_B Z^B_d)
\end{eqnarray}
we see that the box terms from the variations of the scalar and spinor kinetic terms cancel.

Next we concentrate on the first line of the variation of the scalar
kinetic term above and the second term in the variation of the Dirac
operator. To cancel these two terms one needs to introduce the
supercurrent term. Our ansatz for this term reads
\begin{eqnarray}
L_{SC_{1,2}}&=&-Ai(ee_{\alpha}{}^{\mu}e_{\beta}{}^{\nu})\bar\chi_{\mu}^{AB}\gamma^{\beta}\gamma^{\alpha}\Psi_{Aa}
(\tilde D_{\nu}\bar Z^a_B-\tfrac{i}{2}\hat A\bar\chi_{\nu
BC}\Psi^{Ca})\cr
&-&Ai(ee_{\alpha}{}^{\mu}e_{\beta}{}^{\nu})\bar\chi_{\mu
AB}\gamma^{\beta}\gamma^{\alpha}\Psi^{Aa} (\tilde D_{\nu}
Z^B_a-\tfrac{i}{2}\hat A\bar\chi_{\nu}^{ BC}\Psi_{Ca})\,,
\end{eqnarray}
where the index $1$ refers to the first term in the two brackets and
$2$ to the $\hat A$  terms. Terms of the kind we are here seeking to
cancel arise if we vary the two spinors in $L_{SC_{1}}$. Varying
$\chi$ gives
\begin{eqnarray}
-Aie\tilde D_{\mu}\bar\ep
_g^{AB}\gamma^{\nu}\gamma^{\mu}\Psi_{Aa}\tilde D_{\nu}\bar Z_B^a +
c.c.
\end{eqnarray}
while the variation of $\Psi$ produces the result
\begin{eqnarray}
-Aie\bar\chi_{\mu}^{AB}\gamma^{\nu}\gamma^{\mu}\gamma^{\rho}\tilde
D_{\rho}Z^D_a\ep_{mDA} \tilde D_{\nu}\bar Z_B^a+c.c.\,.
\end{eqnarray}

Using the duality flip and the identity
\begin{equation}
\gamma^{\nu}\gamma^{\mu}\gamma^{\rho}=\tfrac{1}{e}\ep^{\nu\mu\rho}+2(g^{\mu(\nu}\gamma^{\rho)}
-\tfrac{1}{2}g^{\nu\rho}\gamma^{\mu})
\end{equation}
we find that demanding cancelation gives two conditions on the matter and gravity supersymmetry parameters:
\begin{equation}
2\ep_g=A\ep_m,\,\,\,\ep_m=A\ep_g,\,\,\,\rightarrow
A=\pm\sqrt{2},\,\,\,\ep_g=\pm\tfrac{1}{\sqrt{2}}\ep_m.
\end{equation}

After these cancellations the remaining $\tilde D^2$ terms are
\begin{equation}
iA(\bar\chi_{\mu}^{AB}\ep_{mAD}-\bar\chi_{\mu
AD}\ep_m^{AB})\ep^{\mu\nu\rho}\tilde D_{\nu}Z^D_a \tilde
D_{\rho}\bar Z^a_B\,,
\end{equation}
which forces us to add a $\chi^2$ term to Lagrangian, namely
\begin{equation}
L_{A'}=iA'\ep^{\mu\nu\rho}\bar\chi_{\mu}^{AC}\chi_{\nu BC}Z^a_A\tilde D_{\rho}\bar Z^B_a+c.c.\,,
\end{equation}
where $\bar\chi\chi$ is automatically in the representation ${\bf 15}$ of $SU(4)$ so the derivative can only
be integrated by parts onto the other scalar field (reality of this term then follows from the
duality flip property). The variation gives, after some integrations by parts,
\begin{eqnarray}
\delta L_{A'}&&=-2iA'(\bar\ep_g^{AC}f_{BC}^{\rho}-\bar f^{\rho AC}\ep_{g BC})(Z^B_a
\tilde D_{\rho}\bar Z^a_A)\cr
&&+iA'\ep^{\mu\nu\rho}(\bar\ep_g^{AC}\chi_{\mu BC}-\bar \chi^{AC}_{\mu}\ep_{g BC})
(\tilde D_{\nu}Z^B_a\tilde D_{\rho}\bar Z^a_A+\tfrac{1}{2}Z^B_a\tilde F_{\nu\rho }{}^a{}_b
\bar Z^b_A+\tfrac{1}{2}Z^B_a
G_{\nu\rho}{}_A{}^D\bar Z_D^a)\cr
&&-A'\ep^{\mu\nu\rho}\bar\chi_{\mu}^{AC}\chi_{\nu BC}(\bar\ep_{m}^{BD}\Psi_{Da}
(\tilde D_{\rho}\bar Z^a_A)+Z^B_a(\tilde D_{\rho}\bar\ep_{mAD})\Psi^{Da}+Z^B_a\bar\ep_{mAD}
(\tilde D_{\rho}\Psi^{Da})).\cr
&&+c.c.\,.
\end{eqnarray}
Thus we find that,  provided $A'=1$, the term $\tilde D_{\nu}Z^B_a\tilde D_{\rho}\bar Z^a_A$
and its complex conjugate cancel the
same terms previously obtained   in the variation of the supercurrent.

This leaves us with the following $\tilde D^2$ terms
\begin{eqnarray}
-2iA'(\bar\ep_g^{AC}f_{BC}^{\rho}-\bar\ep_{gBC}f^{\rho AC})Z_a^B\tilde D_{\rho}\bar Z_A^a+c.c.\,.
\end{eqnarray}

If we now consider the terms obtained by performing the gravitational variation of
the R-symmetry gauge field in the Klein-Gordon term, we find
\begin{eqnarray}
\delta L_1\vert_{\delta B_{\mu}\vert_{grav}}=-i(\bar f^{\nu
AC}\gamma^{\mu} \gamma_{\nu}\ep_{gBC}-\bar
f^{\nu}_{BC}\gamma^{\mu}\gamma_{\nu}\ep_g^{AC}) (Z_a^B(\tilde
D_{\mu}\bar Z_A^a)- (\tilde D_{\mu}Z_a^B)\bar Z_A^a),
\end{eqnarray}
which is entirely a contribution to the $\tilde D^2$ terms. Using the fact  that $A'=1$
found above, this expression can be combined with the
one in the previous paragraph leaving the following $\tilde D^2$ terms in
$\delta L$
\begin{eqnarray}
i(\bar f^{\nu AC}\gamma_{\nu} \gamma^{\mu}\ep_{gBC}-\bar
f^{\nu}_{BC}\gamma_{\nu}\gamma^{\mu}\ep_g^{AC}) (Z_a^B(\tilde
D_{\mu}\bar Z_A^a)- (\tilde D_{\mu}Z_a^B)\bar Z_A^a).
\end{eqnarray}

The next term to be added is
\begin{eqnarray}
L_{A''}=iA''(\bar f^{ AB}\cdot\gamma\Psi_{Aa}\bar Z_B^a+\bar
f_{AB}\cdot\gamma\Psi^{Aa}Z^B_a).
\end{eqnarray}
If we concentrate on the $\tilde D^2$ terms we get two such from the variation of $\chi$ in $f$ and of $\Psi$.
The former gives
\begin{eqnarray}
&&\tfrac{i}{2}A''\ep^{\mu\nu\rho}\tilde D_{\nu}\tilde
D_{\rho}\bar\ep_g^{AB} \gamma_{\mu}\Psi_{Aa}\bar Z^a_B+c.c.\cr
&&=\tfrac{i}{4}A''\ep^{\mu\nu\rho}(-\tfrac{1}{4}\tilde
R_{\nu\rho\alpha\beta}\bar\ep_g^{AB}
\gamma^{\alpha\beta}+2G_{\nu\rho}{}^{[A}{}_C\bar\ep_g^{\vert C\vert
B]})\gamma_{\mu}\Psi_{Aa}\bar Z^a_B+c.c.\,,
\end{eqnarray}
while the latter generates the expression
\begin{eqnarray}
&&iA''\bar f^{BA}_{\nu}\cdot\gamma^{\nu}\gamma^{\mu}(\tilde D_{\mu}
Z_d^C\ep_{mCB}- i\hat A\bar\chi_{\mu}^{CD}\Psi_{Dd}\ep_{mCB}\cr
&&+f^{ab}{}_{cd}Z^C_aZ^D_b\bar Z_B^c\ep_{mCD}
-f^{ab}{}_{cd}Z^D_aZ^C_b\bar Z_C^c\ep_{mDB})\bar Z^d_A+c.c,
\end{eqnarray}
where the first term is a $\tilde D^2$ term, which for $A''=\mp\sqrt{2}$
exactly cancels the traceless part of the previous expression above
leaving just the trace part:
\begin{eqnarray}
&&-\tfrac{iA''}{4}(\bar
f^{AB}\cdot\gamma\gamma^{\mu}\ep_{mAB})(\tilde D_{\mu}(Z\bar Z))\cr
&&=-\tfrac{iA''}{4}(\bar\ep_{gAB}f^{\mu
AB}-\ep^{\mu\sigma\rho}\bar\ep_{gAB}\gamma_{\sigma}f_{\rho}^{AB})(\tilde
D_{\mu}(Z\bar Z)).
\end{eqnarray}

The next term we need is
\begin{eqnarray}
L_{RZ^2}=-\tfrac{e}{8}\tilde R \vert Z\vert^2
\end{eqnarray}
which has the following variation
\begin{eqnarray}
\delta\tilde L_{RZ^2}&=&\tfrac{i}{4e}\vert Z\vert^2\tilde
R_{\mu,\nu}^{**}
\bar\ep_g^{AB}\gamma^{\nu}\chi^{\mu}_{AB}-\tfrac{i}{4e}\tilde R^{**}
(\bar\ep^{AB}_m\Psi_{Aa}\bar Z^a_B+\bar\ep_{mAB}\Psi^{Aa}Z^B_a)\cr
&&+\tfrac{i}{2e}\ep^{\mu\nu\rho}(\tilde D_{\mu} \vert Z\vert^2+
K_{\sigma\mu}{}^{\sigma}\vert Z\vert^2)\bar\ep_g^{AB}\gamma_{\nu}
f_{\rho AB}\cr
&&+\tfrac{i}{2e}\epsilon^{\mu\nu\rho}K_{\mu\nu}{}^{\sigma}\vert
Z\vert^2(\bar\ep_g^{AB}\gamma_{\sigma} f_{\rho
AB}-\tfrac{1}{2}g_{\sigma\rho}\bar\ep_g^{AB}\gamma\cdot f_{AB}),
\end{eqnarray}
where we used the fact that
\begin{eqnarray}
\delta \tilde \omega^*_{\nu}{}^{\alpha\beta}=\tfrac{2i}{e^2}\ep^{\alpha\beta\rho}
(\bar\ep^{AB}_g\gamma_{\nu} f_{\rho AB}-\tfrac{1}{2}g_{\nu\rho}\bar\ep^{AB}_g\gamma_{\sigma} f^{\sigma}_{ AB}),
\end{eqnarray}
and that the double dual of the Ricci tensor with torsion
\begin{eqnarray}
\tilde R_{\nu,\mu}^{**}=\tilde R_{\mu,\nu}-\tfrac{1}{2}g_{\mu\nu}\tilde R,
\end{eqnarray}
where one should note the order of the indices.

Thus we find that two new cancellations occur between the second and
the third terms in this expression and the corresponding ones
proportional to $A''$ above. This leaves three terms at the $\tilde D^2$
level to discuss. One is proportional to the R-symmetry gauge field
and generates an additional contribution to the variation of the
gauge field. The other two are
\begin{eqnarray}
\tfrac{i}{4e}\vert Z\vert^2\tilde R_{\mu,\nu}^{**}
\bar\ep_g^{AB}\gamma^{\nu}\chi^{\mu}_{AB}-\tfrac{i}{4}A''(\bar\ep_{g}^{AB}f^{\mu}_{
AB})(\tilde D_{\mu}(Z\bar Z)).
\end{eqnarray}

The final term we need to add in order to demonstrate that all $\tilde D^2$ terms cancel in $\delta L$ is the
fermionic analog of $L_{R\vert Z\vert^2}$, namely
\begin{eqnarray}
L_{Z^2f\chi}=iA'''\vert Z \vert^2\bar f^{\mu}_{AB}\chi_{\mu}^{AB}
\end{eqnarray}
whose variation reads
\begin{eqnarray}
\delta L_{Z^2f\chi}&=&A'''(\bar\ep_{AB}\Psi^{Aa}Z^B_a+\bar\ep^{AB}\Psi_{Aa}\bar Z^a_B)
\bar  f^{\mu}\chi_{\mu}^{AB}-iA'''(\tilde D_{\mu} \vert Z\vert^2)\bar f^{\mu}_{AB}\ep^{AB}_g\cr
&&-\tfrac{i}{2}A'''\vert Z\vert^2(\tilde R^{**\mu, \nu}\bar \ep_{gAB}\gamma_{\nu}\chi^{AB}_{\mu}-
2\ep^{\mu\nu\rho}G_{\mu\nu}{}^A{}_C\bar\ep_g^{CB}\chi_{\rho AB})\cr
&&+\tfrac{i}{2}A'''\vert Z\vert^2\ep^{\mu\nu\rho}(-\tfrac{1}{4}\delta \tilde \omega_{\nu\alpha\beta}
\bar\chi_{\rho AB}\gamma^{\alpha\beta}\chi^{AB}_{\mu}+2\delta B_{\nu }{}^A{}_C\bar\chi_{\rho}^{CB}\chi_{\mu AB}).
\end{eqnarray}
Thus we see that the last $\tilde D^2$ terms in the previous paragraph are canceled by choosing
$A'''=\tfrac{1}{2}$ and the theory is hence supersymmetric at this order and above in covariant
derivarives. Note that in the two-star curvature term only the symmetric part is $\tilde D^2$.

To cancel the $G_{\mu\nu}$-terms and $F_{\mu\nu}$-terms we obtained above, it is
necessary to add new  terms to the  variations of the gauge fields $A_{\mu}$ and $B_{\mu}$,
chosen such that their Chern-Simons terms give exactly the same  $G_{\mu\nu}$-terms
and $F_{\mu\nu}$-terms but with the opposite sign.

Before analysing the terms with less than two derivatives we summarize what we have found so far:
\begin{eqnarray}
L\vert_{\tilde D^3, \tilde D^2}&=&L_{sugra}^{conf}+L_{BLG}^{cov}
+iA(ee_{\alpha}{}^{\mu}e_{\beta}{}^{\nu})\bar\chi_{\mu}^{AB}\gamma^{\beta}\gamma^{\alpha}\Psi_{Aa}
(\tilde D_{\nu}\bar Z^a_B-\tfrac{i}{2}\hat A\bar\chi_{\nu
BC}\Psi^{Ca})+c.c.\cr
&&+iA'\ep^{\mu\nu\rho}\bar\chi_{\mu}^{AC}\chi_{\nu BC}Z^a_A\tilde
D_{\rho}\bar Z^B_a+c.c.\cr &&+iA''(\bar f^{\mu
AB}\gamma_{\mu}\Psi_{Aa}\bar Z_B^a+\bar
f^{\mu}_{AB}\gamma_{\mu}\Psi^{Aa}Z^B_a)\cr &&-\tfrac{e}{8}\tilde R
\vert Z\vert^2+iA'''\vert Z \vert^2\bar f^{\mu}_{AB}\chi_{\mu}^{AB}
\end{eqnarray}
is supersymmetric to second and third order  in derivatives under the following transformation rules
\begin{eqnarray}
\delta e_{\mu}{}^{\alpha}&=&i\bar\ep_{g AB}\gamma^{\alpha}\chi_{\mu}^{AB} ,\cr
\delta \chi_{\mu}^{AB}&=&i\tilde
D_{\mu}\ep_{g}^{AB} ,\cr
\delta B_{\mu ~B}^{~A}&=&i(\bar f^{\nu AC}\gamma_{\mu}\gamma_{\nu}\ep_{g BC}-
\bar f^{\nu}_{BC}\gamma_{\mu}\gamma_{\nu}\ep_{g}^{AC} )\cr
&&+\tfrac{i}{4}(\bar\ep_{m BD}\gamma_{\mu}\Psi^{a(D}
Z^{A)}_a-\bar\ep_{m}^{AD}\gamma_{\mu}\Psi_{a(D} \bar Z_{B)}^a)
%-\tfrac{i}{4}(\bar\ep_{m BD}\gamma_{\mu}\Psi^{a[D}Z_a^{A]}
%-\bar\ep^{AD}_m\gamma_{\mu}\Psi_{a[D}\bar
%Z_{B]}^a)
\cr
&&-\tfrac{i}{4}(\bar\ep^{AC}_g\chi_{\mu DC}-\bar\ep_{g
DC}\chi^{AC}_{\mu})Z^D_a\bar Z^a_B-
\tfrac{i}{4}(\bar\ep^{DC}_g\chi_{\mu BC}-\bar\ep_{g
BC}\chi^{DC}_{\mu})Z^A_a\bar Z^a_D-trace\cr &&+
\tfrac{i}{8}(\bar\ep^{AD}_g\chi_{\mu BD}-\bar\ep_{g
BD}\chi^{AD}_{\mu})\vert Z\vert^2 ,\cr
 \delta Z^A_a
&=&i\bar\ep_m^{AB}\Psi_{Ba}, \cr
\delta \Psi_{Bd}&=&\gamma^\mu\ep_{m
AB}(\tilde D_\mu Z^A{}_d -i\hat A\bar\chi_{\mu}^{AD}\Psi_{Dd})\cr
 &+&
f^{ab}{}_{cd} Z^C{}_a Z^D{}_b \bar Z_B{}^c
 \epsilon_{m CD}- f^{ab}{}_{cd} Z^A{}_a Z^C{}_b \bar Z_C{}^c \epsilon_{m AB} ,\cr
\delta A_{\mu ~b}^{~a}&=&-i(\bar\ep_{mAB}\gamma_{\mu}\Psi^{Aa}Z^B_b-\bar\ep_m^{AB}
\gamma_{\mu}\Psi_{Ab}\bar Z_B^a)\cr
&&-2i(\bar\ep_g^{AD}\chi_{\mu BD}-\bar\ep_{gBD}\chi_{\mu}^{AD})Z^B_b\bar Z^a_A\,,
\end{eqnarray}
provided the parameters introduced in this section are given the
values
\begin{eqnarray}
\ep_g=\pm\tfrac{1}{\sqrt{2}}\ep_m,\,\,\,A=\pm\sqrt{2},\,\,\,A'=1,\,\,\,A''=\mp\sqrt{2},\,\,\,A'''=\tfrac{1}{2}.
\end{eqnarray}

\subsection{Comments on the results at this stage}

We now turn to the hatted parameter $\hat A $. Having found the
relation between the supersymmetry parameters in the ABJM and
supergravity theories, we can also determine $\hat A $ by requiring
that the variation of $\Psi $ be supercovariant, which gives $\hat
A= \pm\sqrt{2}$. This result will be confirmed in the next section.
There we will also discover that supersymmetry does not demand that
the $\chi\chi ZDZ$ term be supercovariantized which is welcome
because that would mean terms of the type $\chi^3 \Psi Z$. In fact,
from  section two we know that the Lagrangian does not contain any
terms at all with more than two explicit  $\chi$ fields.

Recall also that we concluded that the R-symmetry gauge field
$B_{\mu ~B}^{~A}$ is traceless when checking the local supersymmetry
in the pure supergravity sector. This property must be implemented
also after coupling it to matter. Inspecting the transformation rule
of $\delta B_{\mu ~B}^{~A}$ above we see that this property is
indeed satisfied also when the new terms are included.

Finally we would like to comment on the the abelian gauge field that
is written out explicitly in the Lagrangian in section two. That
this field provides an extra freedom at the order $(\tilde
D_{\mu})^2$ can be seen as follows. Introducing a charge $q$ in the
covariant derivative means that (suppressing R-symmetry indices)
\begin{equation}
[\tilde D_{\mu},\tilde D_{\nu}]Z^a=\tilde F_{\mu\nu}{}^a{}_b Z^b+q
F_{\mu\nu} Z^a.
\end{equation}
The cancelation at this point in the analysis then works as follows.
The relevant terms are
\begin{equation}
\delta L\vert_{D^2}=\epsilon^{\mu\nu\rho}(\tilde F_{\mu\nu}{}^a{}_b
Z^b+q F_{\mu\nu} Z^a)J_{\rho a}+\epsilon^{\mu\nu\rho}(\delta
A_{\mu}{}^a{}_b \tilde F_{\nu\rho}{}^b{}_a +\delta C_{\mu}
F_{\nu\rho}),
\end{equation}
which vanishes provided
\begin{eqnarray}
\delta A_{\mu}{}^a{}_b&=&-Z^aJ_{\mu b}\,,\cr
 \delta C_{\mu}&=&-qZ^aJ_{\mu a},\label{deltaAZJ}
 \end{eqnarray}
that is, for any value of the charge $q$ even though the structure
of $J_{\mu a}$ is dictated by the theory. This fact will be made use
of in the next section.

\section{Cancelation of all terms in $\delta L$ with one covariant derivative}

In this section we continue the program of constructing a
supersymmetric Lagrangian by considering the cancelation of all
terms in $\delta L$ that are of first order in the covariant
derivative. As we will see below this will force us to introduce a
number of new terms in the Lagrangian as well as to add further
terms to the transformation rules presented at the end of the
previous section.

Considering only the field content, there are six different kinds of
terms in $\delta L$ containing one derivative, two of these are
bilinear in fermions, three are quartic and one is of sixth order in
fermionic variables (including the supersymmetry parameter). Some
structures come with different $\gamma$ content and either with or
without a structure constant which makes the list of independent
terms to check a bit longer. We consider the cancelation of these terms in the order of
increasing number of fermions. This will not fix the Lagrangian
completely although the final form of the transformation rules will
be determined. In the next section we will extend the analysis
to terms in $\delta L$ which have two fermions and no derivatives. The information then
obtained will be enough to
give the final answer also for the Lagrangian.

To be more precise this part of the analysis will force us to add
new terms to the supersymmetry transformation rules, that is, terms
over and above those specified at the end of the last section. In
particular we will need in $\delta \Psi$ new $Z^3$ terms without a
structure constant:
\begin{eqnarray}
\delta \Psi_{Bd}\vert_{new}=\tfrac{1}{4}Z^C_cZ^D_d\bar
Z^c_B\ep_{CD}+\tfrac{1}{16}\vert
Z\vert^2Z^A_d\ep_{AB}.\label{deltaPsiZcube}
\end{eqnarray}
It will also become clear from the calculations below that the 
underlying ABJM matter theory must be extended by an extra $U(1)$
gauge field as we have already mentioned in previous sections. In the
final Lagrangian presented in section two the Chern-Simons term for
this gauge field (denoted $C_{\mu}$) was given explicitly. There its transformation rule
was also presented and in this section we  will see how these features of the theory arise. 

Note that in the headings below $\cdot f$ refers to the fact that
the term contains a structure constant $f^{ab}{}_{cd}$ and that the
derivative can be acting on any of the fields although it is
generally written as acting on the scalars.

\subsection{Terms of second order in fermionic variables}

\subsubsection{$e(\bar\epsilon\gamma^{\mu}\Psi)\tilde D_{\mu}Z^3\cdot f$}

This calculation is needed already in the ungauged ABJM case. The
new feature here is a remaining term where the derivative acts on
the supersymmetry parameter. Such terms are easily canceled by
adding new terms in the Lagrangian containing a $\chi_{\mu}$ that
when varied gives rise to the same kind of unwanted terms in $\delta
L$ but with opposite sign. These new terms are here of the form
$e\chi\Psi Z^3\cdot f$ and appear  in the Lagrangian in section two
as the terms on the line (\ref{hatted:B}).
%directly after the terms
%that were derived in the previous section.

\subsubsection{ $e(\bar\ep\gamma^{\mu}\Psi)\tilde D_{\mu}Z^3$}

The terms considered here are similar to the ones  just analyzed apart
from the important fact that they do not contain a structure
constant. Such terms arise due to the presence of the new $\ep\Psi
Z$ terms  (without structure constants) that we found were necessary
to add to $\delta B_{\mu}$ in the previous section. These new terms  will, however,  
create problems
 when used in the variation of the Klein-Gordon term. Our approach to deal with this will contain
 additional modifications of the transformation rules together
 with new Yukawa-like terms without structure constants in the Lagrangian.

 Thus we add to the Lagrangian the five possible structures that can be built out of
 two $\Psi$ and two $Z$ fields
 without using a structure constant. These terms are then varied under
 $\delta \Psi\vert_{DZ}$. To get this to work it turns out necessary to first
 modify $\delta\Psi$ by adding to it two $Z^3$ terms without structure
 constants (see (\ref{deltaPsiZcube}))
and consider the variation of the Dirac term and, secondly, to
introduce an extra $U(1)$ gauge field that plays a
special role. To this end we give the corresponding gauge field the
following transformation rule
 \begin{eqnarray}
 \delta C_{\mu}\vert_{\psi}=-iq(\bar\epsilon_{AB}\gamma_{\mu}\Psi^{Aa}Z^B_a-
 \bar\epsilon^{AB}\gamma_{\mu}\Psi_{Aa}\bar Z_B^a),
\end{eqnarray}
and the ABJM matter fields charge $q=\pm\tfrac{1}{4}$. When adding
the corresponding variation of the Klein-Gordon term we find that
there remains only a term with the derivative acting on the
parameter. This last term we can cancel as usual by adding $\chi\Psi Z^3$
terms without structure constants to the Lagrangian.

\subsubsection{$e\ep^{\mu\nu\rho}(\bar\ep\gamma_{\mu}\chi_{\nu})\tilde D_{\rho}Z^4\cdot f$ and
$e(\bar\ep\chi^{\mu})\tilde D_{\mu}Z^4\cdot f$}

These terms arise from the variation of the $\delta \tilde A_{\mu}$
in the Klein-Gordon term, the $\delta\Psi\vert_{Z^3\cdot f}$
variation in the first part of the supercurrent term and the
$\delta\Psi\vert_{DZ}$ in the new $\chi\Psi Z^3 \cdot f$ term in the
Lagrangian. Adding these we find that the terms without
$\ep^{\mu\nu\rho}$ cancel directly while the ones with an epsilon
tensor do not. However, also the "f-terms" $\bar f
\cdot \gamma \Psi Z$ varied under $\delta \Psi\vert_{Z^3 \cdot f}$
contributes to the epsilon terms and when added leave only a
$\tilde D_{\mu}\ep$ term which is canceled by adding an
$\epsilon^{\mu\nu\rho}\bar\chi_{\mu}\gamma_{\nu}\chi_{\rho}Z^4$ term
to $L$.

\subsubsection{$e\ep^{\mu\nu\rho}(\bar\ep\gamma_{\mu}\chi_{\nu})\tilde D_{\rho}Z^4$ and
$e(\bar\ep\chi^{\mu})\tilde D_{\mu}Z^4$}

By varying the first part of the supercurrent under
$\delta\Psi\vert_{Z^3}$ and the $\chi\Psi Z^3$ term under
$\delta\Psi\vert_{DZ}$ we get contributions to both structures
considered here. The epsilon tensor terms are canceled by the new
$\delta\Psi\vert_{Z^3}$ variation (\ref{deltaPsiZcube}) of the $\bar
f\Psi Z$ term and the $\chi\chi Z^4$ terms without structure
constants. To cancel the non-epsilon terms we need to vary the
Klein-Gordon term with respect to $B_{\mu}$ to find that once again
we seem to need a special $U(1)$ gauge field that varies into $\chi$
according to
\begin{eqnarray}
 \delta C_{\mu}\vert_{\chi}=-2iq(\bar\epsilon_g^{AD}\chi_{\mu BD}-\bar\epsilon_{gBD}\chi_{\mu}^{AD})Z^B_a\bar
    Z^a_A\,,
\end{eqnarray}
again leading to the value  $q=\pm\tfrac{1}{4}$. Note that to get the same value for $q$
we have normalized this variation
and the previous one $\delta C_{\mu}\vert_{\Psi}$ in the same way as for the corresponding terms in the
variation of the non-abelian gauge field $\delta\tilde A_{\mu}$.

\subsection{Terms quartic in fermionic variables}

\subsubsection{$e\bar\ep\chi \tilde D \Psi^2$}

Terms of this kind come from the Dirac term by varying the dreibein,
the supercovariant spin connection, the R-symmetry gauge field, and
the ABJM spinor field itself. To these four contributions we add the
terms obtained by performing a $\delta Z$ variation in the first
part of the supercurrent and in the so called f-term, namely $\bar
f\cdot\gamma\Psi\bar Z+c.c.$. What remains to be canceled after
these terms are added together are terms with the derivative acting
on the susy parameter. These final terms are exactly canceled by the
variation of the second part of the supercurrent provided we write
the ABJM Dirac term in a manifestly real way after gauging, i.e., by
replacing
\begin{eqnarray}
- i \bar \Psi^{Aa} \gamma^\mu D_\mu  \Psi_{Aa}\rightarrow -
\tfrac{ie}{2} \bar \Psi^{Aa} \gamma^\mu  \tilde D_\mu  \Psi_{Aa}-
\tfrac{ie}{2} \bar \Psi_{Aa} \gamma^\mu \tilde D_\mu  \Psi^{Aa},
\end{eqnarray}
since this will mean that an otherwise required $\chi^2\Psi^2$ term
is automatically accounted for.

\subsubsection{$e\bar\ep\chi  \chi\Psi \tilde D Z$}

The analysis here fixes the coefficient in the Lagrangian of the
term that would supercovariantize the $\chi\chi Z\tilde DZ$ term, namely
$e\chi^3\Psi Z$. We find that this term has a vanishing coefficient
and hence no term in the Lagrangian is of higher order than two in
explicit $\chi$ fields.

The calculation goes as follows. We add the contributions from the
$\delta Z$ in $e \chi^2 Z\tilde DZ$, $\delta \Psi$ in the Dirac term
and $e\chi^2\Psi^2$, $\delta B_{\mu}$ variations of the Chern-Simons
term for the gravitino, $\delta B_{\mu}$, $\delta \Psi$ and $\delta
e$ of the $\chi\Psi \tilde DZ$ term, the $\delta\omega_{\mu}$,
$\delta B_{\mu}$, $\delta\Psi$ and $\delta e$ of the term $f\Psi Z$
and finally $\delta Z$ of the $f\chi Z^2$ term. The result is
\begin{eqnarray}
-\tfrac{1}{2}(\bar f^{\mu AB}\gamma^{\nu}\chi_{\mu
AB})(\bar\Psi^{Ca}\gamma_{\nu}\epsilon_{CD})Z^D_a+c.c.\,.
\end{eqnarray}
However, this is exactly canceled by a term used already in the
previous chapter on cancelation of $(\tilde D_{\mu})^2$ terms, namely the
Riemann tensor term that arises in the $\delta \chi$ variation of
$L_{A''}$. As we know from the supergravity analysis  the double dual of the Riemann
tensor is a second rank tensor whose  symmetric piece is second order in
derivatives while the antisymmetric part contains only one derivative. This latter tensor
is just, after dualization,
the triple dual whose variation gives   the above term with opposite sign.

\subsubsection{$e\bar\ep\chi \chi^2 \tilde D \vert Z\vert^2$}

Terms of this kind arise from the Chern-Simons term for the
gravitino field, the $e\chi^2Z\tilde DZ$ term, and the two terms $e\tilde R
Z^2$ and $e\bar f \chi Z^2$. From the fact that these cancel we conclude that
the term that would
supercovariantize the $e\chi\chi Z\tilde DZ$ term, i.e. $e\chi^2 Z
\chi\Psi$, has zero coefficient confirming the result
obtained in the previous subsection. This calculation is similar to
the one just above but makes instead use of the Riemann tensor
coming from the variation of the term $L_{RZ^2}$.

\subsection{Terms of sixth order in fermions}

These terms are all of the form $e(\ep\chi)\tilde D \chi^4$ and do
not arise explicitly in the variation of any of the terms in the
Lagrangian. Thus all such terms are hidden in the covariant derivatives
and therefore  automatically dealt with when canceling the derivative terms.

\subsection{Comments on the use of the $U(1)$ gauge field}

Here we continue the discussion of the extra abelian vector field that was 
started
at the end of the previous section. We saw there that it was
possible to give the ABJM matter fields a charge $q$ under the corresponding $U(1)$ 
gauge symmetry, and furthermore that this
charge was not determined  by the cancelation of terms of order
$(\tilde D_{\mu})^2$ in $\delta L$. However, as we have seen above, and now
explain in more detail, the value of this charge is fixed by the
order $\tilde D_{\mu}$ analysis performed in this section.

The terms relevant for this discussion are first of all the terms
that remain after canceling the (non-ABJM gauge field) variations at first order
in derivatives, that is,
\begin{equation}
\delta L\vert_{remaining}=(Z^a J_{\mu
b})(K^{\mu})^b{}_a+\tfrac{1}{16} Z^a J_{\mu a}(K^{\mu})^a{}_a,
\end{equation}
 where $(K^{\mu})^b{}_a$ is a fixed expression, and secondly the total matter gauge
field variation
\begin{equation}
\delta L\vert_{\tilde D_{\mu}}=\delta \tilde A_{\mu}{}^a{}_b
(K^{\mu})^b{}_a+q \delta C_{\mu} (K^{\mu})^a{}_a.
\end{equation}

Combined with (\ref{deltaAZJ}) these variations cancel each other
provided $q^2=\tfrac{1}{16}$. The corresponding new transformation
law for the abelian gauge field is thus found to be
\begin{eqnarray}
 \delta C_{\mu}&=-iq(\bar\epsilon_{AB}\gamma_{\mu}\Psi^{Aa}Z^B_a-
 \bar\epsilon^{AB}\gamma_{\mu}\Psi_{Aa}\bar Z_B^a)\cr
    &-2iq(\bar\epsilon_g^{AD}\chi_{\mu BD}-\bar\epsilon_{gBD}\chi_{\mu}^{AD})Z^B_a\bar
    Z^a_A.
\end{eqnarray}

\section{Cancelation of bifermion non-derivative terms in $\delta L$}

In this section we add all terms  in the Lagrangian that do not
give any derivative contributions to its variation , i.e., different kinds of $eZ^6$
terms. By demanding cancelation  of
all non-derivative two-fermion terms in $\delta L$  the coefficients  in $L$
of these $eZ^6$ terms are determined which finalizes  the structure also of the Lagrangian.

The relevant terms that must cancel arise from the old and new Yukawa terms
with $\Psi$ varied into the ABJM $Z^3$ term with an $f$ and the new
terms of this kind without an $f$. Certain combinations of these expressions
then cancel the contributions coming from varying $Z$ in the various
kinds of potential terms as will now be explained.

\subsection{$e(\bar\ep\Psi)Z^5\cdot f^2$}

These $f^2$ terms are known to  cancel already  in the original ABJM
computation, which is valid also here since no new contributions of
this kind arise in the coupled theory.

\subsection{$e(\bar\ep\Psi)Z^5\cdot f$}

These terms are similar to the previous ones but with only one
structure constant. They arise from several sources: first from the
$\delta\Psi\vert_{Z^3}$ variation of the ABJM Yukawa term and
secondly from $\delta\Psi\vert_{Z^3\cdot f}$ variation of the five
new Yukawa like terms without structure constant. When adding these
up the result can be seen to cancel the variation of  $Z$ in the new
potential term with one $f$.

\subsection{$e(\bar\ep\Psi)Z^5$}

In the same fashion as for the previous cancelation these terms
arise from the new Yukawa like terms without $f$ by varying $\Psi$
into $Z^3$ without $f$. Some of these terms eliminate each other,
while the remaining terms cancel the variation of $Z$ in the new $f$-free
potential term.

\subsection{$e(\bar\ep\gamma\cdot\chi)Z^6\cdot f^2$}

This kind of $f^2$ term comes from the $\delta\Psi\vert_{Z^3}\cdot
f$ variation of the $\chi\Psi Z^3\cdot f$ and must cancel the
dreibein variation of the $Z^6 \cdot f^2$, i.e.,  the original ABJM
potential term in $L$, which it does.

\subsection{$e(\bar\ep\gamma\cdot\chi)Z^6\cdot f$}

Terms with one  structure constant arise from the
$\delta\Psi\vert_{Z^3}$ variation of $\chi\Psi Z^3\cdot f$ in $L$
and from $\delta\Psi\vert_{Z^3\cdot f}$ variation of $\chi\Psi Z^3$.
After cycling of the indices in one of the terms they cancel exactly
the dreibein variation of the $Z^6 \cdot f$ term in $L$.

\subsection{$e(\bar\ep\gamma\cdot\chi)Z^6$}

These terms (without structure constant) arise from the
$\delta\Psi\vert_{Z^3}$ variation of the $\chi\Psi Z^3$ terms in the
action and from the variation of the dreibein in the $Z^6$ potential
without structure constants. After cycling the $SU(4)$ indices in
one of the terms, and using the self-duality relation, all
terms can be seen to cancel.

\section{Conclusions}\label{concl}

In this paper we have coupled a general ABJM theory to the corresponding
conformal supergravity theory constructed in section three of this
paper. The proof of supersymmetry of the coupled theory has been
carried through for all terms in $\delta L$ with three, two and one
derivative, together with all terms without derivatives that are
bilinear in fermionic variables (including the susy parameter). This
has been described in detail in the previous sections.

We will now discuss the remaining eight (non-derivative) terms in
$\delta L$.  Note that at this point in the analysis, i.e., before
checking these last terms in $\delta L$, the Lagrangian itself is in
fact completely determined which is true also for the transformation
rules. This follows from the fact that the only terms in the
Lagrangian that do not generate any derivatives when varied are the
pure $eZ^6$ terms. To determine their coefficients it is then
sufficient to consider the cancelation of all terms in $\delta L$
with two fermions. Concerning the transformation rules any term
added at the non-derivative stage would alter parts of the previous
calculations involving terms with derivatives and
invalidate it.

Thus we conclude that the Lagrangian and the transformation rules
presented in section two of this paper constitute the complete
answer. The last terms in $\delta L$ that must be analyzed in order
to finalize the proof of supersymmetry are the following
(non-derivative) ones, ordered in decreasing number of $\chi$
fields,
\begin{eqnarray}
\bar\ep \chi \chi^6, \,\,\bar\ep \chi \chi^4
Z^2,\,\,\bar\ep\Psi\chi^4Z,\,\,\bar\ep \chi \chi^2\Psi^2,\,\,\bar\ep
\chi \chi^2 Z^4,\,\,\bar\ep\Psi\chi^2Z^3,\,\, \bar\ep \chi \Psi^2
Z^2,\,\,\bar\ep\Psi\Psi^2Z.
\end{eqnarray}
 Of these the first one is part of the pure
supergravity calculation, while the second and third are part
 of the covariant derivatives in the coupled theory since
the torsion terms have been kept throughout the calculation. This
fact also account for the fourth kind of term in the list. However,
explicit terms with this field content arise in addition from
varying, e.g., the dreibein in the Dirac term (plus an integration
by parts) and from the term that supercovariantizes  the
supercurrent term in the Lagrangian. That the coefficient of this
explicit $e\chi^2\Psi^2$ term in the Lagrangian is the correct one
to provide this supercovarintization has been verified by checking
the cancelation of terms in $\delta L$ with one derivative. Of the
remaining terms in the above list also the ABJM terms
$e\bar\ep\Psi\Psi^2Z$ have been verified to cancel. Thus the
analysis includes in particular all terms in the original ABJM
theory. What remains to be done is to check the cancelation of the
fourth, fifth, sixth and seventh expressions in the list above. This
is a rather elaborate calculation and has not yet been done in full
detail.

Note that the last four structures in the list above 
can appear both with and without
a structure constant $f$. Of these we have only, as just mentioned,
checked the last one which is just an ABJM computation when it contains
a structure constant. However, when it does not it is more
interesting since then it makes use of the variation of the $U(1)$
gauge field $C_{\mu}$ in the Dirac term and 
therefore gives additional support for the way this
field  is being used here.

Since in this paper the parameters appearing in the Lagrangian and
transformation rules are determined uniquely and in almost all cases
from at least two separate calculations, we are fairly convinced
that the cancelations that have not been established here will not
alter any of our conclusions. Nevertheless, it would be welcome to
find an independent argument for why the construction in this paper
is correct. Methods that have been used in the past in similar
circumstances are, e.g., constrained gauged superconformal algebras,
superspace, the embedding tensor technique \cite{Bergshoeff:2008bh}
and the construction of the on-shell supersymmetry algebra. Although
the first was used early on to obtain the pure conformal
supergravity theories and the latter three were utilized in the more
recent constructions of non-gravity M2 matter theories with eight
(BLG), six (ABJM) or fewer supersymmetries, none of them seem to
straightforwardly give an argument that would guarantee the
existence of the type of coupled theories we are considering here.
We hope to come back to these issues in a future publication.

It is worth remarking that the scalar potential after gauging
contains, apart from the original ABJM terms with two structure
constants, also terms with one as well as no structure constant (see
the last two lines of the Lagrangian presented in section 2.1). As a
further check of the derivation of these new contributions to the scalar potential
one should verify that theory leads to an acceptable set of physical states.
 Another term that is crucial in
this context is the conformal coupling between the curvature scalar
and two scalar fields that arises in the process of checking
supersymmetry. By giving the scalar field a vacuum expectation value
the theory can be related to the corresponding one for a stack of D2
branes \cite{Mukhi:2008ux}. If we insert the VEV into the potential terms with one or two structure
constants one finds that they do not contribute to the cosmological constant
while the remaining potential terms (without structure constant)
give a non-zero contribution. Using a VEV chosen such that it turns the
$-\tfrac{e}{8}\tilde R \vert Z \vert^2$ term into a correctly
normalized Einstein-Hilbert term, one finds a theory where this
term is accompanied by a gravity Chern-Simons term and a
cosmological constant. This part of the theory is described by the following Lagrangian
\begin{equation}
L=-\tfrac{e}{\kappa^2}(R+\tfrac{2}{\kappa^4})+
\tfrac{1}{2}\epsilon^{\mu\nu\rho}Tr(\omega_{\mu}\partial_{\nu}\omega_{\rho}+
\tfrac{2}{3}\omega_{\mu}\omega_{\nu}\omega_{\rho})\,.
\end{equation}
We note that up to a sign this
Lagrangian (with $\kappa^2=16\pi G$) is the same as the one of Li, Song and Strominger \cite{Li:2008dq} at a chiral
point\footnote{This remains the case also if a level (or dimensionless coupling constant)
is introduced in the conformal gravity sector as discussed below. See \cite{Ni:2010aa}
for further details.}.
The chirality is a welcome result while the  sign may be
problematic in view of the discussion in ref. \cite{Li:2008dq}
about the energy of physical states (black holes) and the central charge of the boundary CFT.

Some final comments are in order. First we note that the rather
simple connection that exists between the $SU(2)\times SU(2)$ ABJM
theory and the BLG theory seems less trivial after coupling these
two theories to conformal supergravity. One complicating factor is
that the topologically gauged ABJM theory seems to rely on the
presence of an extra  $U(1)$ gauge field. The supersymmetry exhancement of
ABJM theories with abelian gauge fields has been discussed in
\cite{Aharony:2008ug, Benna:2008zy, Klebanov:2008vq}. It may also be
of some interest to set  the structure constants to zero eliminating
the non-abelian parts of the ABJM gauge group and consider what
might be a non-trivial new theory for one conformal M2 brane with
six supersymmetries. A slightly more involved case arises if we set
$f^{ab}{}_{cd}=\delta^{ab}_{cd}$ which also solves the fundamental
identity. 

In connection with the abelian gauge field and the charge
$q=\pm\tfrac{1}{4}$ assigned to the matter fields, it may be
interesting to reconsider the normalization of the Chern-Simons term
since the level chosen for this term affects the value of $q$. In
fact, since also gravitational Chern-Simons terms are associated
with levels \cite{Horne:1988jf} the general issue of levels in
topologically gauged BLG \cite{Gran:2008NT} and ABJM theories should
be studied further. Note that if we introduce an independent level
in the supergravity sector, or equivalently a dimensionless
gravitational coupling constant  at the classical level, it should
appear in the Lagrangian and in the transformation rules in such a
way as to make it possible to decouple the gravity and matter
sectors by turning it off.

\acknowledgments

We would like to thank Ulf Gran, Horatiu Nastase and Andrew Strominger for discussions.
The work is partly funded by the Swedish Research Council.

\end{document}